\documentclass[preprint,12pt]{elsarticle}

\usepackage{amssymb}

\journal{Physica A}

\usepackage[hyphens]{url} 
\usepackage{subfig} 
\usepackage{rotating} 
\usepackage{comment}
\usepackage{tikz}
\usepackage{float}
\usepackage[hyphens]{url} 
\usepackage{subfig} 

\usepackage{multicol}
\usepackage[utf8]{inputenc}
\usepackage{csquotes}

\usepackage{caption}
\usepackage{amsmath}
\usepackage{changepage}
\usepackage{courier}
\usepackage{colortbl} 

\usepackage{booktabs} 

\usepackage{float}
\usepackage{placeins} 
\usepackage{geometry} 
\usepackage{verbatim}
\usepackage{xcolor} 
\definecolor{light-gray}{gray}{0.92}
\usepackage{listings} 
\usepackage{listingsutf8}
\usepackage{booktabs} 

\usepackage{tabularx}

\usepackage{graphicx}
\usepackage{pdflscape}
\usepackage{adjustbox}

\graphicspath{{./}}

\usepackage[labelfont=bf]{caption} 

\usepackage[labelsep=period]{caption}

\begin{document}

\begin{frontmatter}




    \title{Impact of memory and bias in kinetic exchange opinion models on random networks}


	 \author[uceff,INCT]{André L. Oestereich\corref{cor1}}

	 \author[uff]{Nuno Crokidakis}

	 \author[eco,INCT,LAMFO]{Daniel O. Cajueiro}

	 \address[uceff]{Unidade Central De Educação Faem Faculdade, Itapiranga/SC, Brazil}

	 \address[uff]{Instituto de F\'isica, Universidade Federal Fluminense, Niter\'oi/RJ, Brazil}

	 \address[INCT]{National Institute of Science and Technology for Complex Systems (INCT-SC), Brazil.}

	 \address[eco]{Department of Economics, University of Brasilia, Campus Universitario Darcy Ribeiro - FACE, Brasilia, DF 70910-900.}

	 \address[LAMFO]{Machine Learning Laboratory in Finance and Organizations (LAMFO), Universidade de Bras\'{i}lia (UnB), 70910-900, Bras\'{i}lia, Brazil.}


	 \cortext[cor1]{Corresponding author: andreoestereich@uceff.edu.br }

	\begin{abstract}
        In this work we consider the effects of memory and bias in kinetic exchange opinion models defined on random Erdos-R\'enyi networks. We propose a model in which agents remember the sign of their last interaction with each one of their pairs. This introduces memory effects in the model, since past interactions can affect future ones. We have also considered the impact of a parameter $q$ that regulates how often an agent changes its interaction to match its opinion, thus introducing bias in the interactions. For high values of $q$ an agent is more likely to start having a negative interaction with an agent of opposing opinion and a positive interaction with an agent of the same opinion. The model is defined on the top of random networks with mean connectivity $\langle k \rangle$. We analyze the impact of both $q$ and $\langle k \rangle$ on the emergence of ordered and disordered states in the population. Our results suggest a rich phenomenology regarding critical phenomena, with the presence of metastable states and a non-monotonic behavior of the order parameter. We show that the fraction of neutral agents in the disordered state decreases as the bias $q$ increases.

	\end{abstract}


\begin{highlights}
\item Memory effects in kinetic exchange opinion models
\item The introduction of memory leads to the emergence of bias in the agents' interactions
\item Presence of metastable states and nonmonotonic effects
\end{highlights}

	\begin{keyword}
		Opinion Dynamics \sep Collective Phenomena
		\newline

	\end{keyword}

\end{frontmatter}

\section{Introduction}

Opinion dynamics is one of the most studied problems in the branch of Sociophysics. Despite the simplicity of the models, they are able to mimic the main ingredients of complex social processes such as decision making, elections and diffusion of fake news/rumors \cite{2012galam,2008galam,2014senC,CROKIDAKIS2022127598}.

Although most opinion dynamics' models do not consider effects of memory, they can be an important components to improve these models. For example, such effects were considered recently in the voter model \cite{PERALTA2020122475} and in the noisy voter model \cite{Peralta_2020}. For the voter model, the authors analyzed the model at mean-field level, and they showed that the functional form of the activation probability (related to the memory) has a crucial impact on how and whether the system reaches consensus or not \cite{PERALTA2020122475}. In the case of the noisy voter model, the authors found a rich critical phenomena which includes continuous phase transitions, discontinuous transitions and tricritical behavior \cite{Peralta_2020}. Despite the fact that discontinuous transitions are less common in the context of social systems, they were found in some recent works \cite{2009castellanoFL,2017sirbuLST,abramiuk2021discontinuous,crokidakis2022simple,OESTEREICH2020109893,e21050521,2017gambaroC}.

Yet considering memory effects in opinion dynamics, a recent work considered such effects on the q-voter model \cite{castellano2009nonlinear,2018jedrzejewskiS}. The authors of \cite{2018jedrzejewskiS} considered the q-voter model with social temperature and memory effects modeled considering the extreme value memory. Extreme value memory states that agents' future choices depend on the peak memory of their past experiences \cite{harris2015random}. Thus, the results showed that due to the memories of the past experiences related to each type of social response, if the level of social temperature is below a critical value, the agents may acquire personal traits.

A collective memory can also emerge in opinion dynamics. The authors of \cite{BOSCHI2020124909} proposed a stochastic model that includes dynamically evolving couplings, which effectively record an exponentially weighted history of co-expressed opinions between any pair of agents in the system. They showed how such mechanism allows a society to develop a collective memory of a previous exposed external information  allowing it to spontaneously retrieve such information in the future when briefly triggered by exposure to that information.
The model was also extended in order to analyze how exposure to strong external information can shape the behaviour of a society. The authors verified that people tend to agree with others if they have a history of predominant mutual agreement. On the other hand, they are be more likely to disagree with others in the case of a history of predominant recent disagreement. This mechanism gives rise to a collective memory effect by which a society can remember past configurations of opinions \cite{BOSCHI2021125799}.

People have a tendency to weigh opinions that agree with theirs in a more positive light than those that oppose their views \cite{1979lordRL,1992griffinT}. This psychological effect is known as confirmation bias. This phenomenon can be found in many different contexts and forms \cite{1998nickerson}. Confirmation bias is specially prevalent in online interactions \cite{2015bessiPVZ,2015zolloNVB}. This tends to lead to the fragmentation and polarization of public opinion \cite{2016vicarioBZP}.

Due to its importance there have been efforts to integrate this bias in opinion dynamics models \cite{2014allahverdyanG,2021liuWCT,2022anagnostopoulosBCP}. More specifically, in \cite{2000deffuantNAW} the authors proposed a model where agents ignore opinions that are too different from their own. Inter-group bias was also shown to lead to polarization for disagreeing groups in a kinetic exchange opinion model \cite{2019oestereichPC}.

In this work we consider another mechanism of memory and confirmation bias in kinetic exchange opinion models \cite{2012biswasCS,2010lallouacheCCC}. In section 2 we present the model and the microscopic rules that govern the dynamics. In section 3 we discuss our analytical and numerical results. Finally, the conclusions are presented in section 4.


\section{Model}

Our model consists of a variant of the three-state BCS (Biswas-Chatterjee-Sen) model \cite{2012biswasCS}. The BCS is part of a class of models called kinetic exchange opinion models, that are based on models of wealth exchange. Specifically, in the BCS each agent carries one opinion $o=+1, -1$ or $0$. The dynamical rule is simple: we choose 2 agents at random, say $i$ and $j$ with respective opinions $o_i$ and $o_j$. They interact through a kinetic rule, and the opinion of agent $i$ is updated according $ o_i = \text{sign}( o_i + \mu_{ij} o_j)$, where $\text{sign}(x)$ is the sign function and $\{\mu_{ij}\}$ are random variables such that $\mu_{ij}=-1 \, (+1)$ with probability $p$ $(1-p)$. The BCS model was defined in a fully-connected population, and it undergoes a continuous phase transition at $p_c=1/4$, that separates an ordered ferromagnetic phase and a disordered paramagnetic one \cite{2012biswasCS}. In three-state kinetic exchange opinion models, an ordered phase is defined the presence of a majority opinion ($o=+1$ or $o=-1$), and the coexistence with neutral individuals ($o=0$). On the other hand, a disordered phase is defined by equal fractions of $o=+1$ and $o=-1$ opinions.

In our formulation of the model, we considered that the  couplings $\{\mu_{ij}\}$ among agents $i$ and $j$ evolve in such a way that agents remember if their last interaction was positive or negative. We also assume that the opinions can have tree discrete values $o \in \{-1, 0, 1\}$.

We initiate the model with $N$ agents connected by an Erdos-R\'enyi random network. We assign to each agent $i$ an uniform random opinion ($o_i$) from the set $\{-1, 0, 1\}$. We set fraction $f_0$ of the connections of the network to be negative and the remaining connections to be positive. We consider the adjacency matrix to be symmetric, which is consistent with an undirected graph \cite{RAQUEL2022127825}. Therefore if agent $i$ interacts negatively with agent $j$ ($\mu_{ij} = -1$), then agent $j$ also interacts negatively with agent $i$ ($\mu_{ji}=-1$).

For each time step we carry out $N$ updates. Each update consists of choosing an agent $i$ and one of its neighbors $j$ at random.  If one of the agents has neutral opinion, then $o_i$ is always updated according to

\begin{equation}
    o_i = \text{sign}( o_i + \mu_{ij} o_j).
    \label{interact}
\end{equation}

\noindent  If the pair of agents have opposing opinions ($o_i o_j = -1$), then, with probability $q$, their connection becomes negative ($\mu_{ij} = \mu_{ji} = -1$) and, with probability $1-q$, the opinion $o_i$ is updated according to Eq.~(\ref{interact}). If the agents have the same opinion ($o_i o_j = 1$), then, with probability $q$, their connection becomes positive  ($\mu_{ij} = \mu_{ji} = +1$) and, with probability $1-q$, opinion $o_i$ is updated according to Eq.~(\ref{interact}). This mechanism introduces memory and bias to the model, since past interactions affect future interactions and the sign of the interactions correlates with their combined opinions.

In the subsequent sections we take the average of the opinions $O= \frac{1}{N} \sum o_i$ to be the global order parameter and $f$ to be the fraction of the negative nonzero elements in the adjacency matrix, i.e.

\begin{equation}
    f = \frac{\text{number of negative connections}}{\text{number of connections}} = \frac{1}{2} \left(1 - \frac{\sum_{ij} \mu_{ij}}{\sum_{ij} |\mu_{ij}|} \right).
\end{equation}

\noindent Also, $\langle k \rangle$ is the average number of connections per node in the network.

\begin{figure}[h!]
    \centering
    \includegraphics[width=0.4\textwidth]{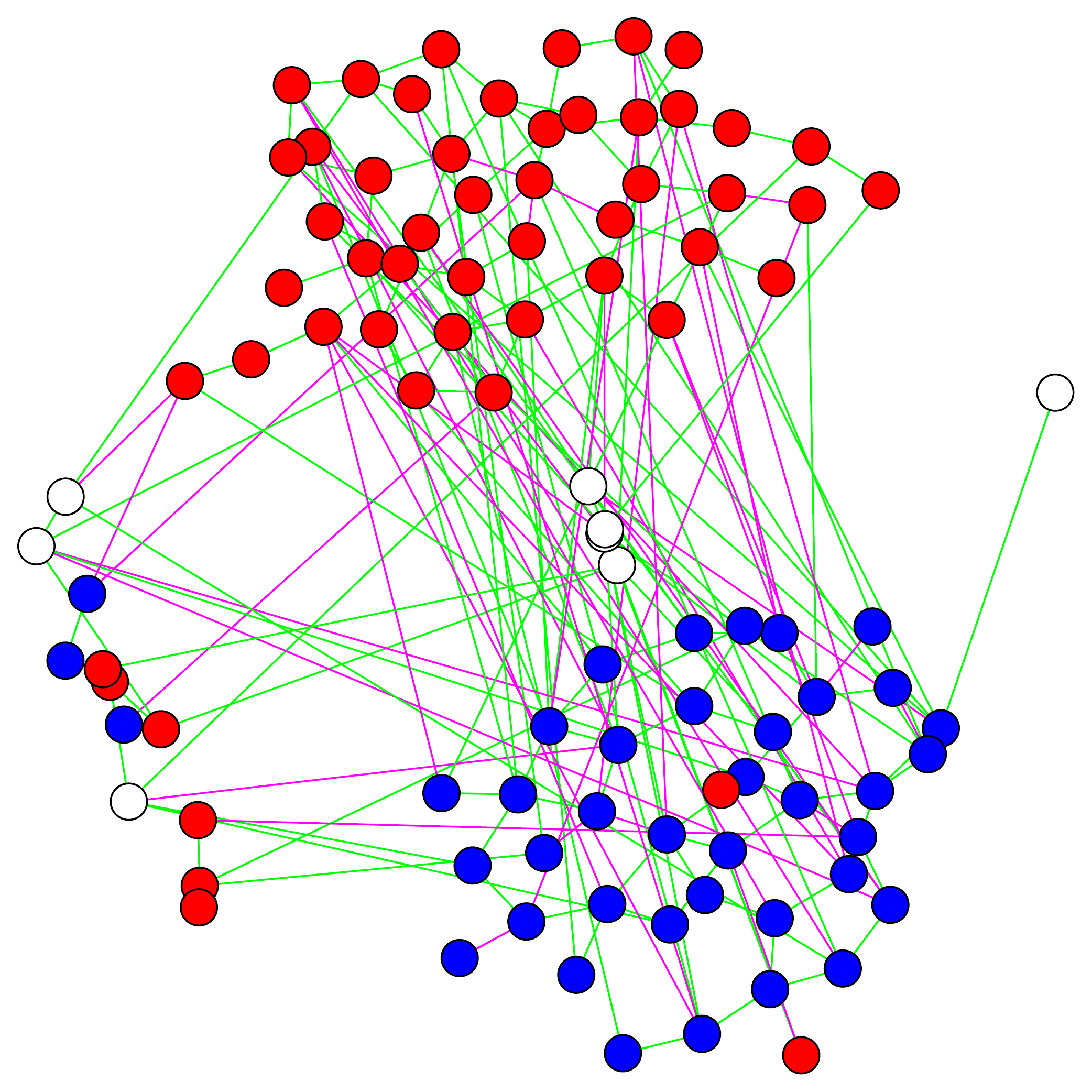}
    \caption{State of the network for $N=100$, $\langle k \rangle = 4, q = 0.3$ and $f_0 = 0.4$. The red circles are agents with negative opinions, the blue ones have positive opinions and the white are the neutrals. The green links represent a positive connection and the magenta ones represent a negative connection.}
    \label{graph}
\end{figure}

\section{Results and Discussion}

In this section we present the results of the Monte Carlo simulations in networks with $N = 10^4$ connected agents. We connect the agents according to random networks \cite{1959erdosR} with varying degrees of connectivity. The results are an average over 50 samples.

An example of stationary state for a small network, $N=100$, is presented in Figure \ref{graph}. The average number of connections between agents with the same or different opinions is not different, but agents with differing opinions tend to have a negative interaction between them. In this figure we can also see that neutral agents tend to have a similar number of positive and negative connections to each of the other opinions. And also, that in this model contrarian individuals can arise naturally. These individuals are mostly negatively connected to individuals of differing opinion.

\begin{figure}[h!]
    \centering
    \subfloat[Various values of $f_0$.]{\includegraphics[width=0.49\textwidth]{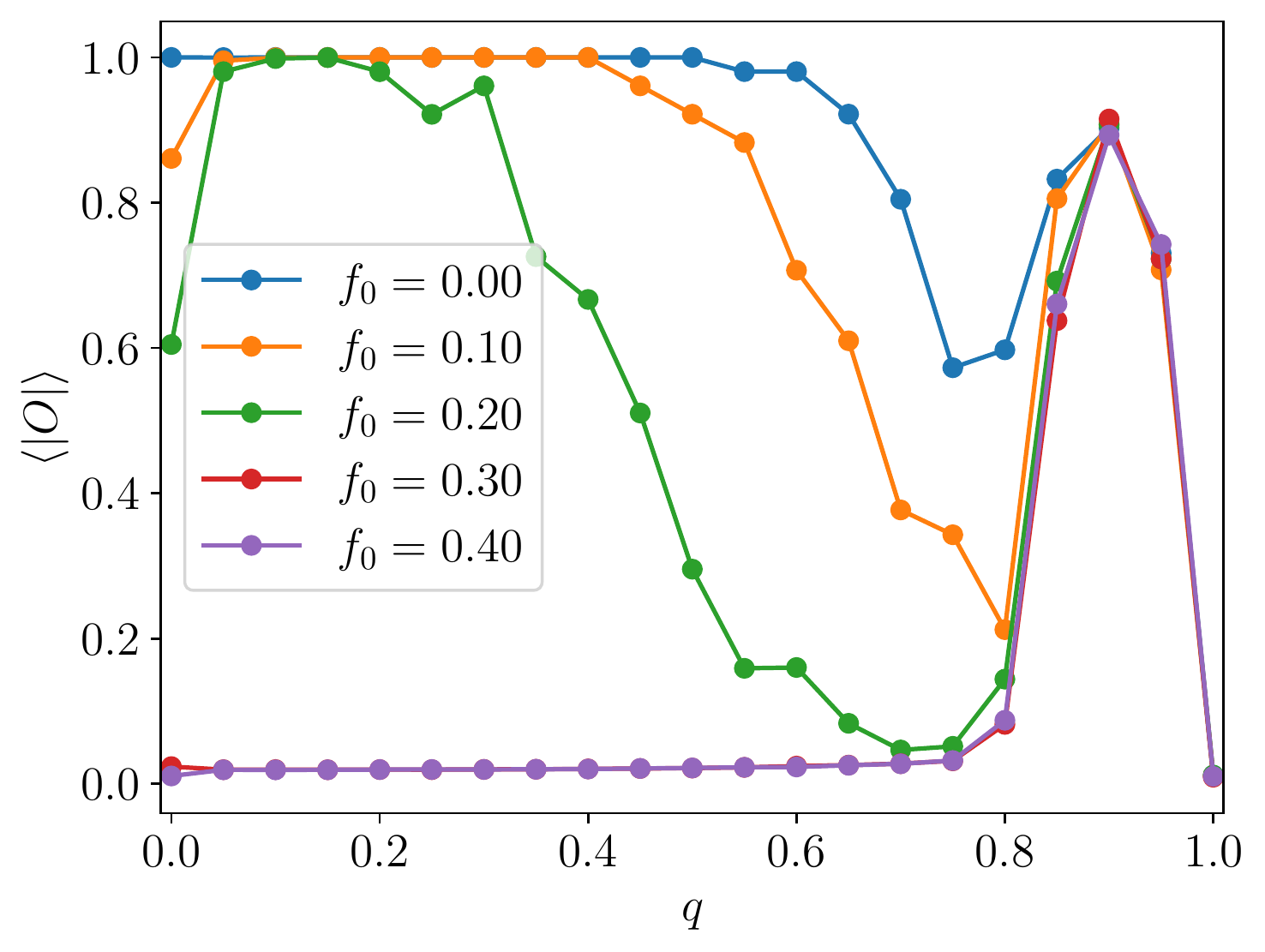}}
    \subfloat[$f_0 = 0.2$.]{\includegraphics[width=0.49\textwidth]{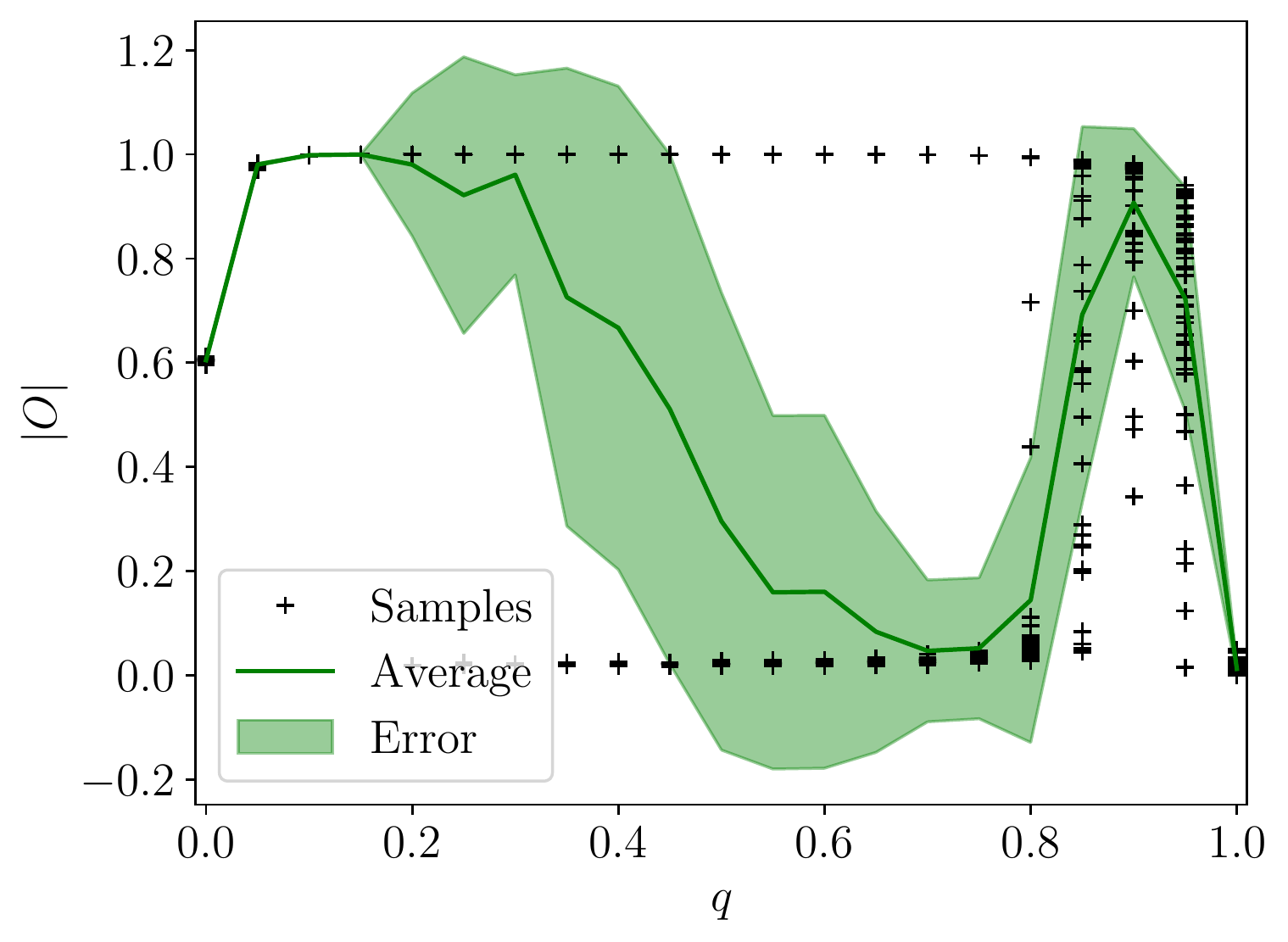}}
    \caption{Stationary states for the order parameter as a function of $q$ for $\langle k \rangle = 50$, different values of $f_0$ in (a) and values for the samples average and standard deviation for $f_0=0.2$ in (b). The increase of $f_0$ reduces the stationary value of the order parameter. This inversely matches very closely with what we observe in Figure \ref{fxp}.}
    \label{OxP}
\end{figure}

Figure \ref{OxP} presents the order parameter of the system as a function of $q$ for different values of $f_0$. For all different initial states the system is non-monotonic. For smaller values of $q$ ($q<0.8$) the system exhibits two metastable states, an ordered one and a disordered one, as made evident in (b). For higher values of $q$ ($q>0.8$) there is a single stationary state and this dependence on the initial conditions of the system disappears.

\begin{figure}[h!]
    \centering
    \includegraphics[width=0.6\textwidth]{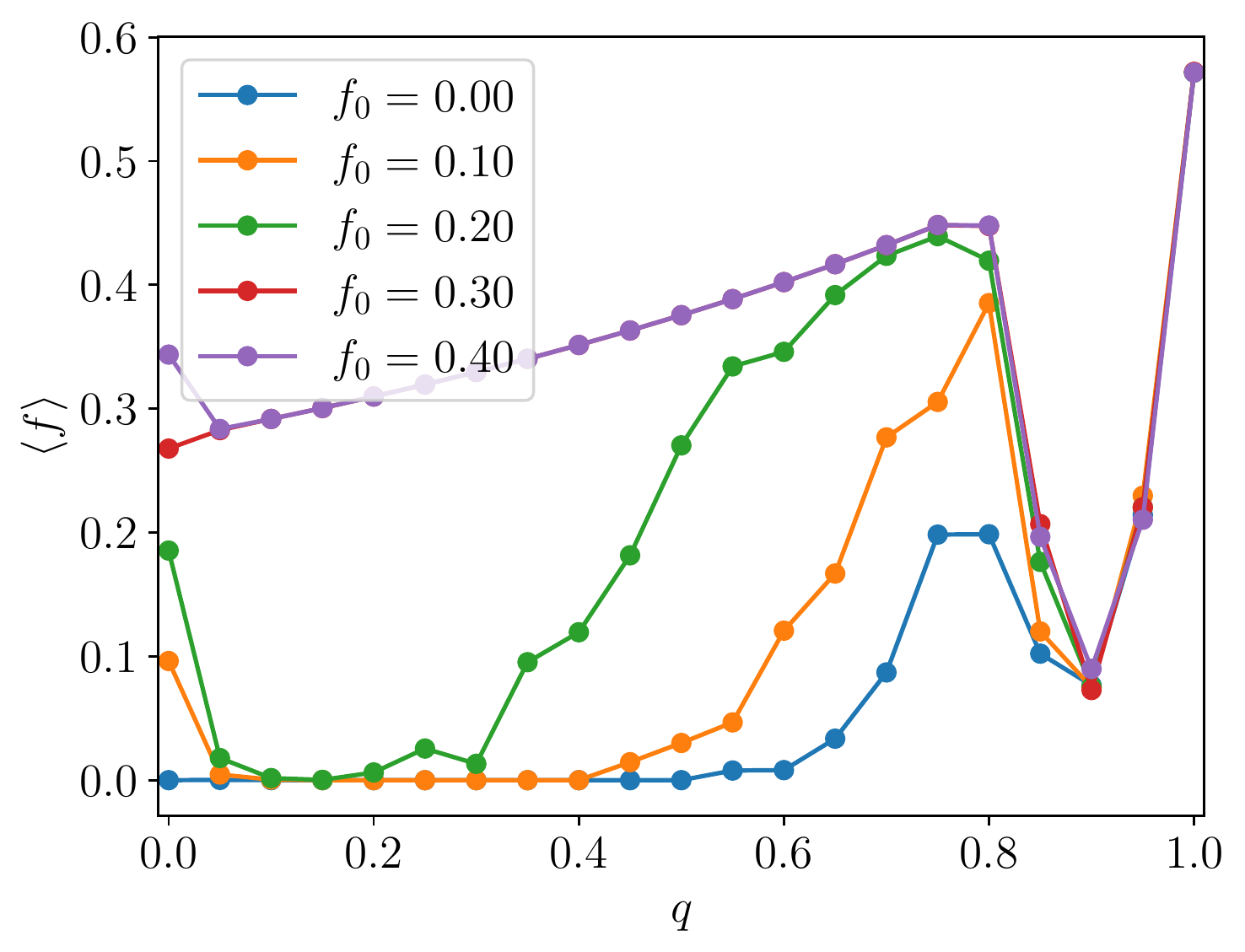}
    \caption{Average value of the stationary states for $f$ as a function of $q$ for $\langle k \rangle = 50$ and different initial values of $f$. Here we observe that by increasing $f_0$ has an effect for $q < 0.8$, maximum values are found for $f_0 = 0.3$. This inversely matches very closely with what we observe in Figure \ref{OxP}.}
    \label{fxp}
\end{figure}

These results are inversely correlated to the stationary values of the fraction of negative connections in the network that are presented in Figure \ref{fxp}. This relation is to be expected since a smaller number of negative connections in the stationary state leads to an ordered state and higher fraction of negative connections leads to a disordered state. Notice that for values of $q \in (0,0.8]$ $f$ reaches a ceiling for each value of $q$, this holds true for $f_0 > 0.4$ as well.

\begin{figure}[h!]
    \centering
    \includegraphics[width=0.6\textwidth]{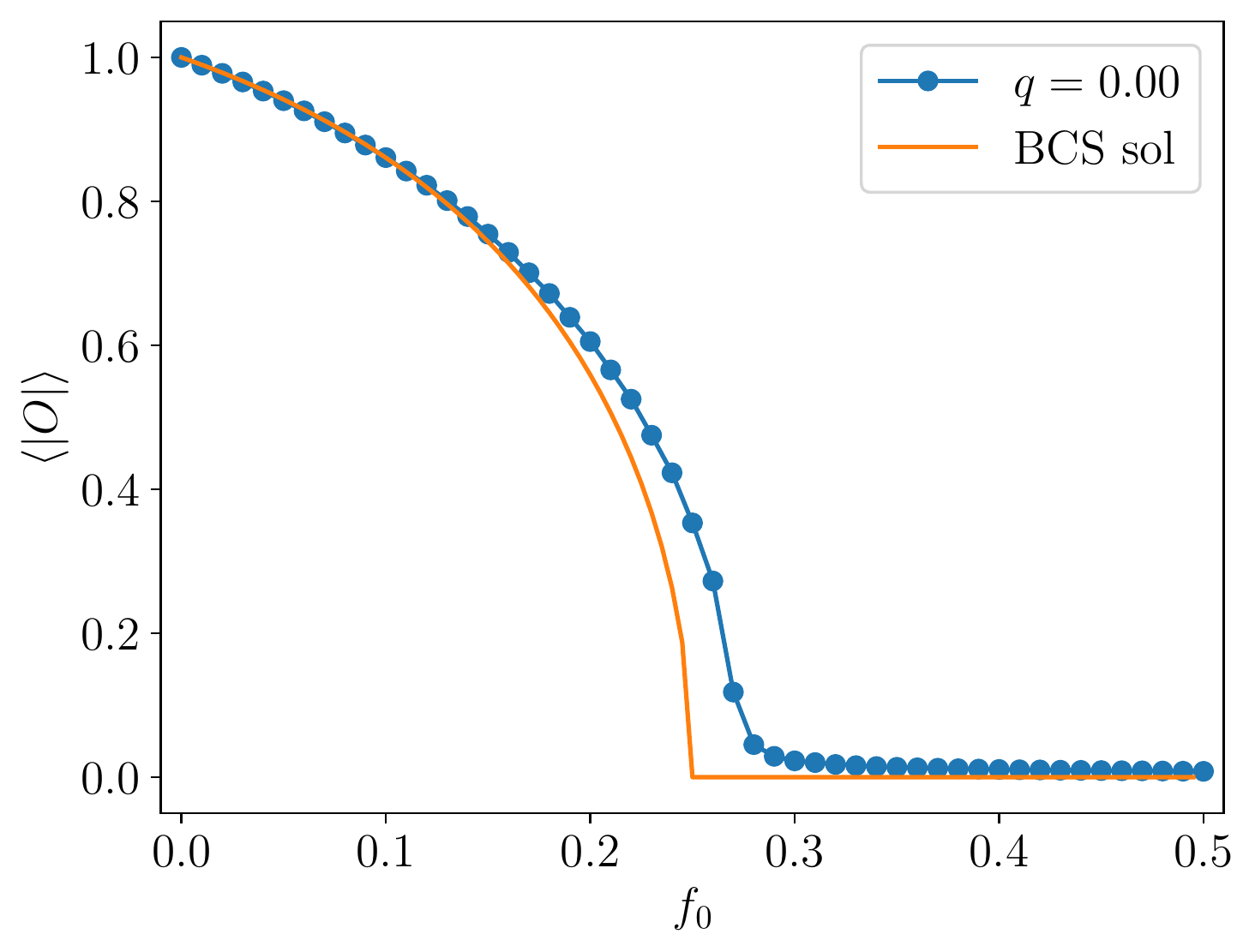}
    \caption{Comparison between the analytical solution of the BCS model and the order parameter for the stationary state for $q=0$ and $\langle k \rangle = 50$. The behaviour is very similar in both cases but in our model the critical value is slightly bigger.}
    \label{OxH}
\end{figure}

In the BCS model when interactions are negative with probability $0.25$ (the critical point of the original BCS model) or higher the system necessarily goes to the disordered state, as can be seen in Figure~\ref{OxH}. In our model $f$ gives the approximate probability of a negative interaction. When $q=0$ the signs of the connections do not change in time, therefore our model with $q=0$ is equivalent to a BCS model with quenched negative interactions. Figure~\ref{OxH} presents a comparison for the results of our model with $q=0$ and the analytical solution for the BCS model, given by $O = \frac{\sqrt{1-4p}}{1-p}$ \cite{2012biswasCS}. We see that because the negative and positive interactions are always between the same individuals the critical point is slightly bigger than the BCS model. It is also apparent, that when the signs of the interactions do not change the results are very similar to the BCS model, therefore the ability of the interactions to change signs in a biased way plays a decisive role in the novelty of the results.

\begin{figure}[h!]
    \centering
    \subfloat[$q=0.1$.]{\includegraphics[width=0.32\textwidth]{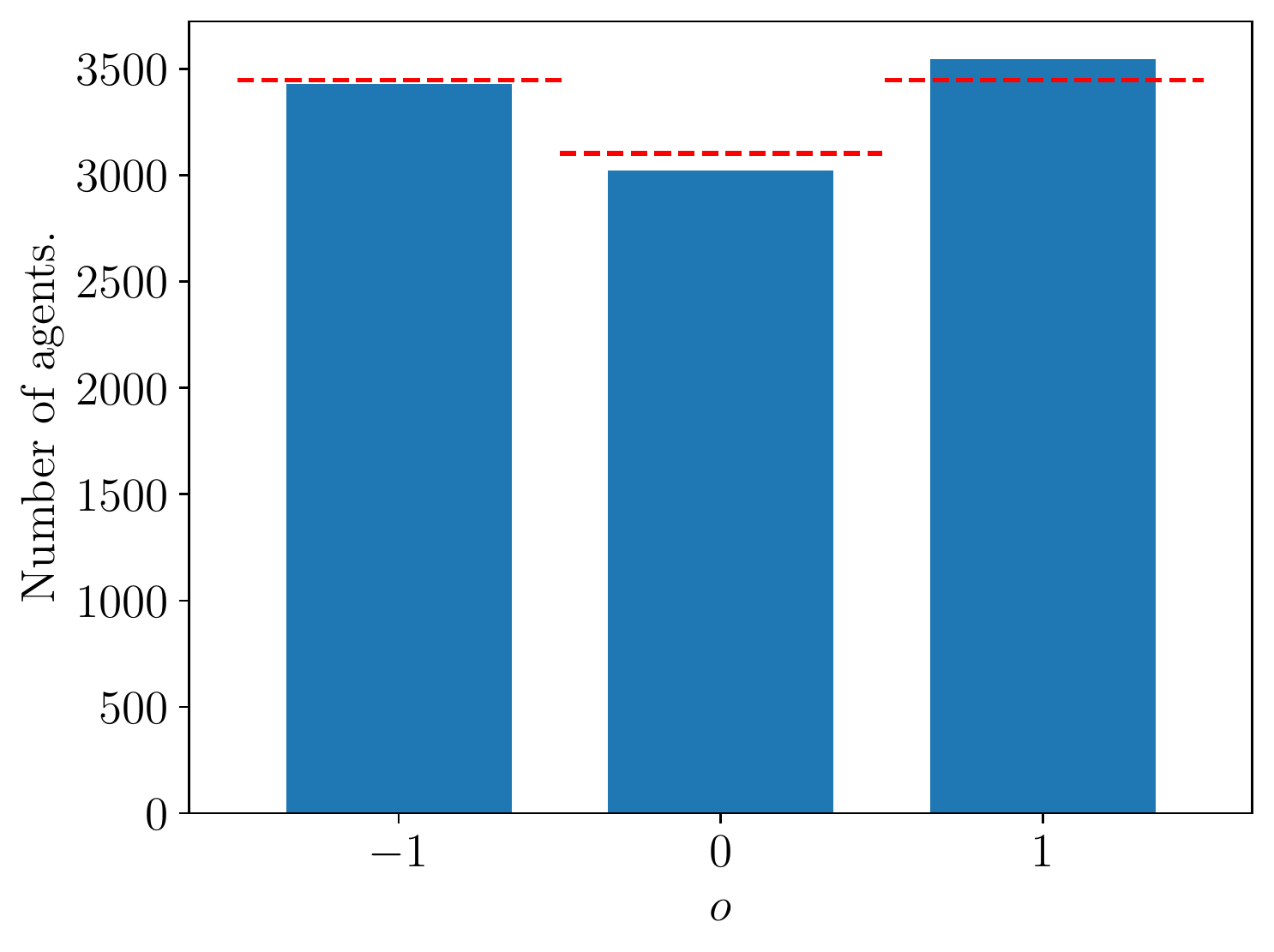}}
    \subfloat[$q=0.5$.]{\includegraphics[width=0.32\textwidth]{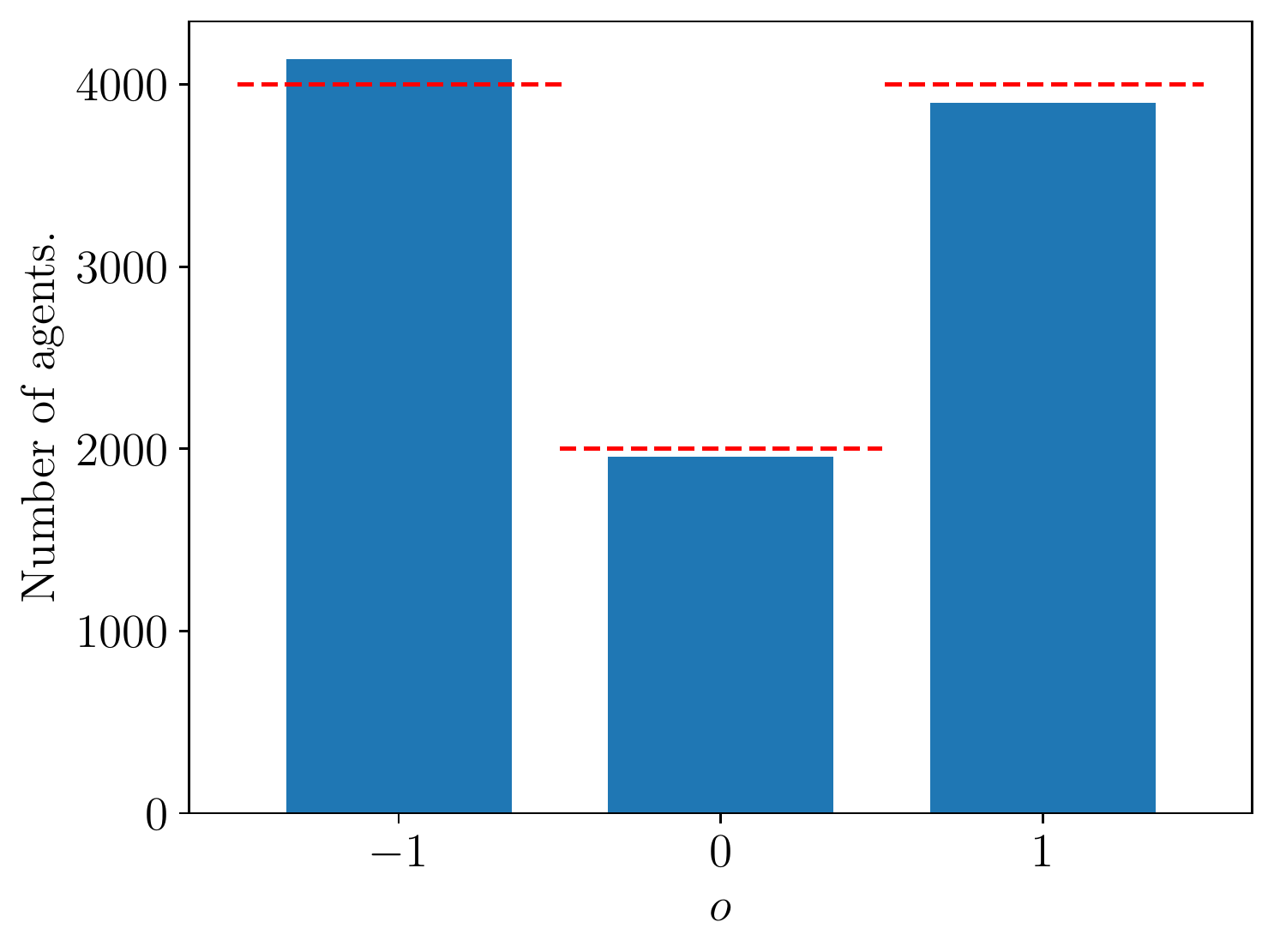}}
    \subfloat[$q=1$.]{\includegraphics[width=0.32\textwidth]{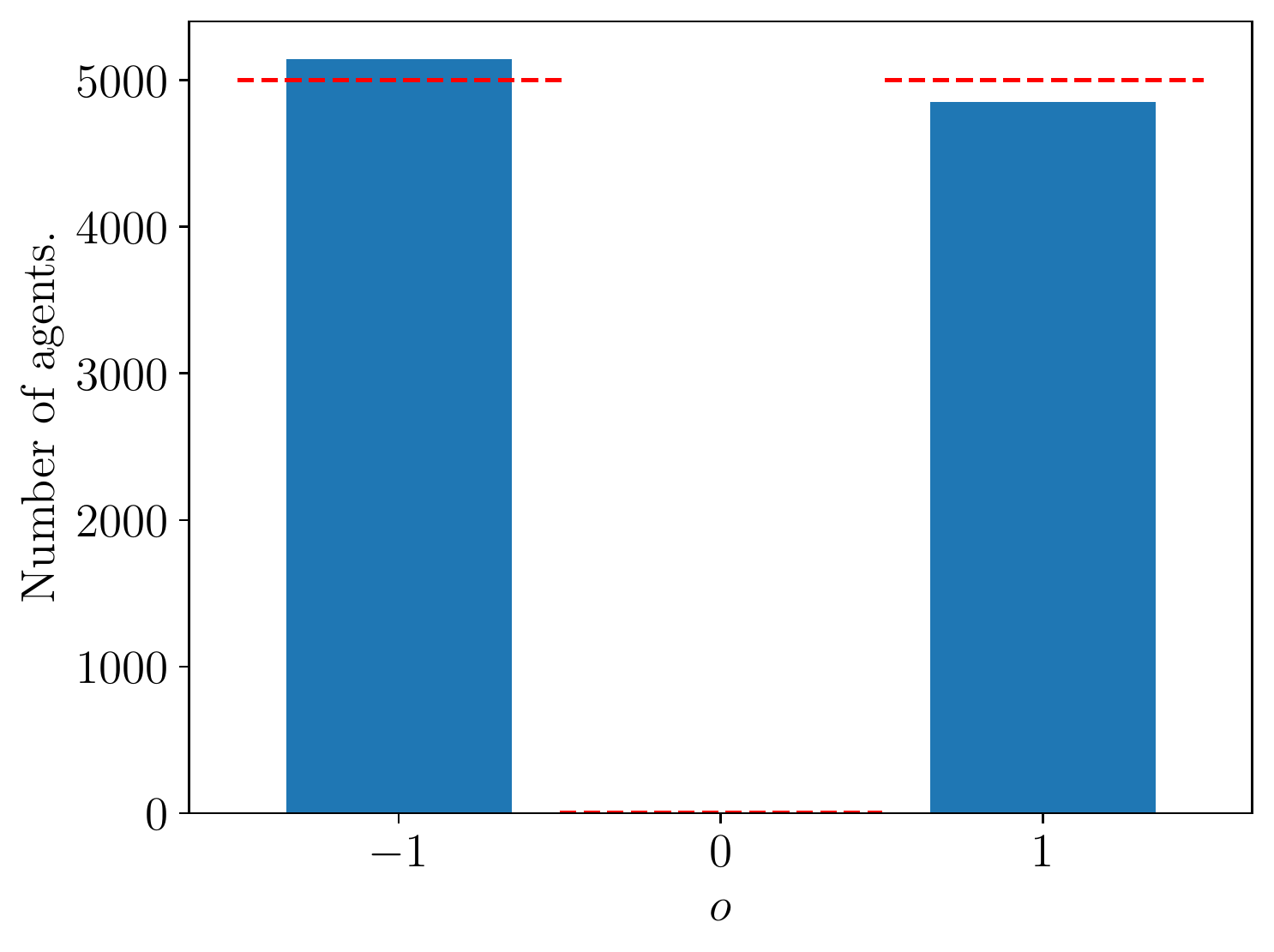}}
    \caption{Histogram of opinions in the disordered state for $f_0 = 0.3$ and $\langle k \rangle = 50$. The fraction of each opinions follows closely the results of the approximated version presented as the red dashed line. As the probability of a connection to change increases the fraction of neutral agents decreases.}
    \label{hist_p}
\end{figure}

From the mean field approximation (see the Appendix) we have that the fraction of neutral opinions in the disordered state is given by

\begin{equation}
    g_0 = \frac{1-q}{3-q}.
    \label{fracs}
\end{equation}

\noindent This is in agreement with the results found in the simulations as can be seen in Figure~\ref{hist_p}. The mean field approximation is not able to capture characteristics of the order parameter for this model since these are strongly correlated with the connectivity of the network and these characteristics are completely ignored in the approximation.

The fraction of neutral opinions is reduced by increases in $q$ since this increases the probability of an agent changing their connection instead of their opinion. This means that, for $q=1$ it is not possible for the system to find consensus, because in order to change one's opinion it first has to become neutral and for $q=1$ there are no neutral agents.

\begin{figure*}[h]
    \centering
    \includegraphics[width=0.6\textwidth]{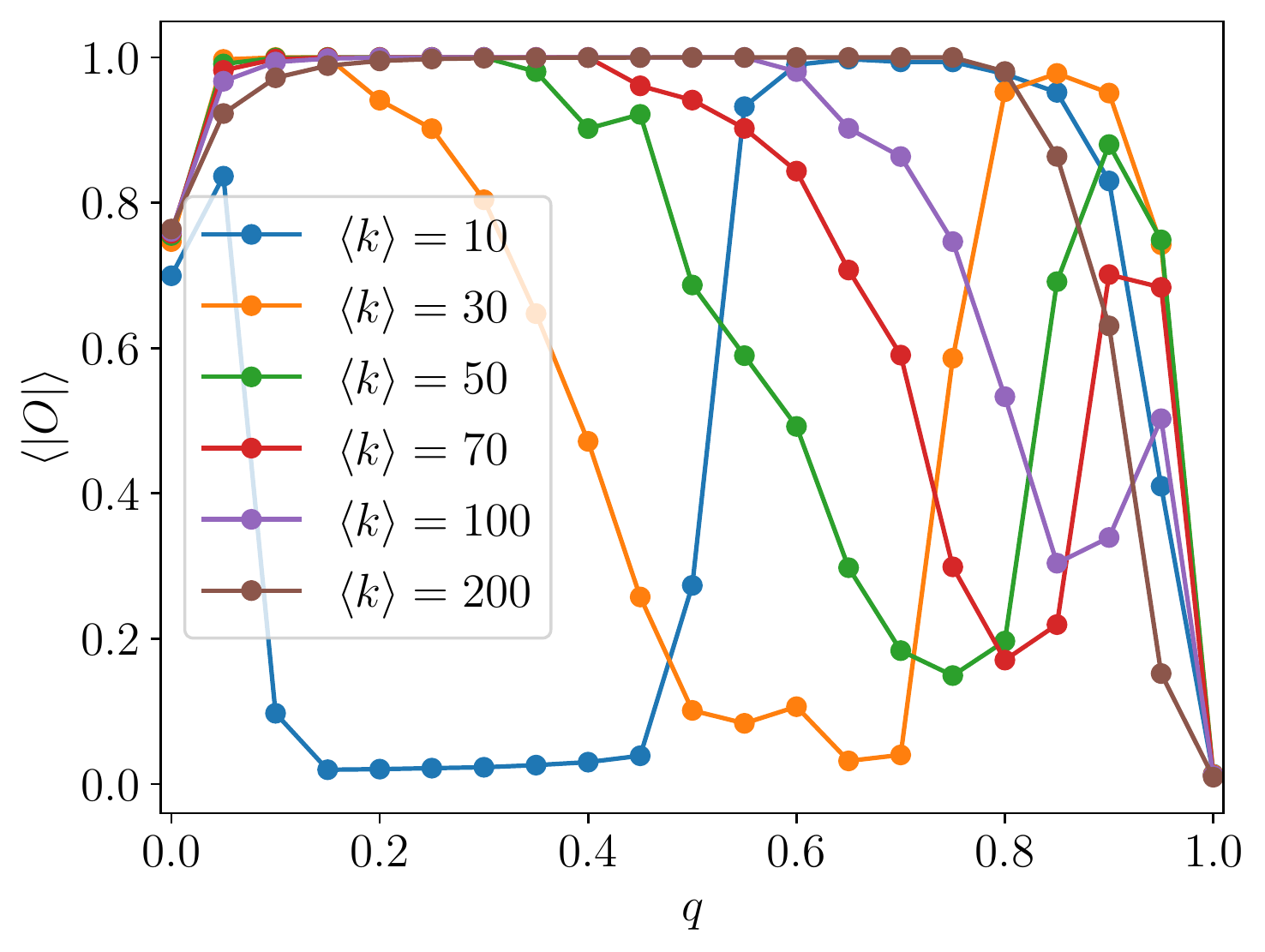}
    \caption{Influence of the connectivity of the network in the stationary state of the average opinion for $f_0 = 0.15$. Here we can see that the size of the network has a considerable effect in the outcome of the dynamic, and this is why the approximation of the model is not very accurate.}
    \label{OxP_ks}
\end{figure*}

The role of $\langle k \rangle$ is shown in Figure~\ref{OxP_ks} where order parameter as a function of $q$ is presented. In this graph we see that for higher values of $\langle k \rangle$ the range of values of $q$ that have a disordered phase becomes smaller as the values of $q$ increases. Thus, increases in $\langle k \rangle$ tend to increase the region of values of $q$ that lead to the ordered phase.

\begin{figure*}[h]
    \centering
    \includegraphics[width=0.6\textwidth]{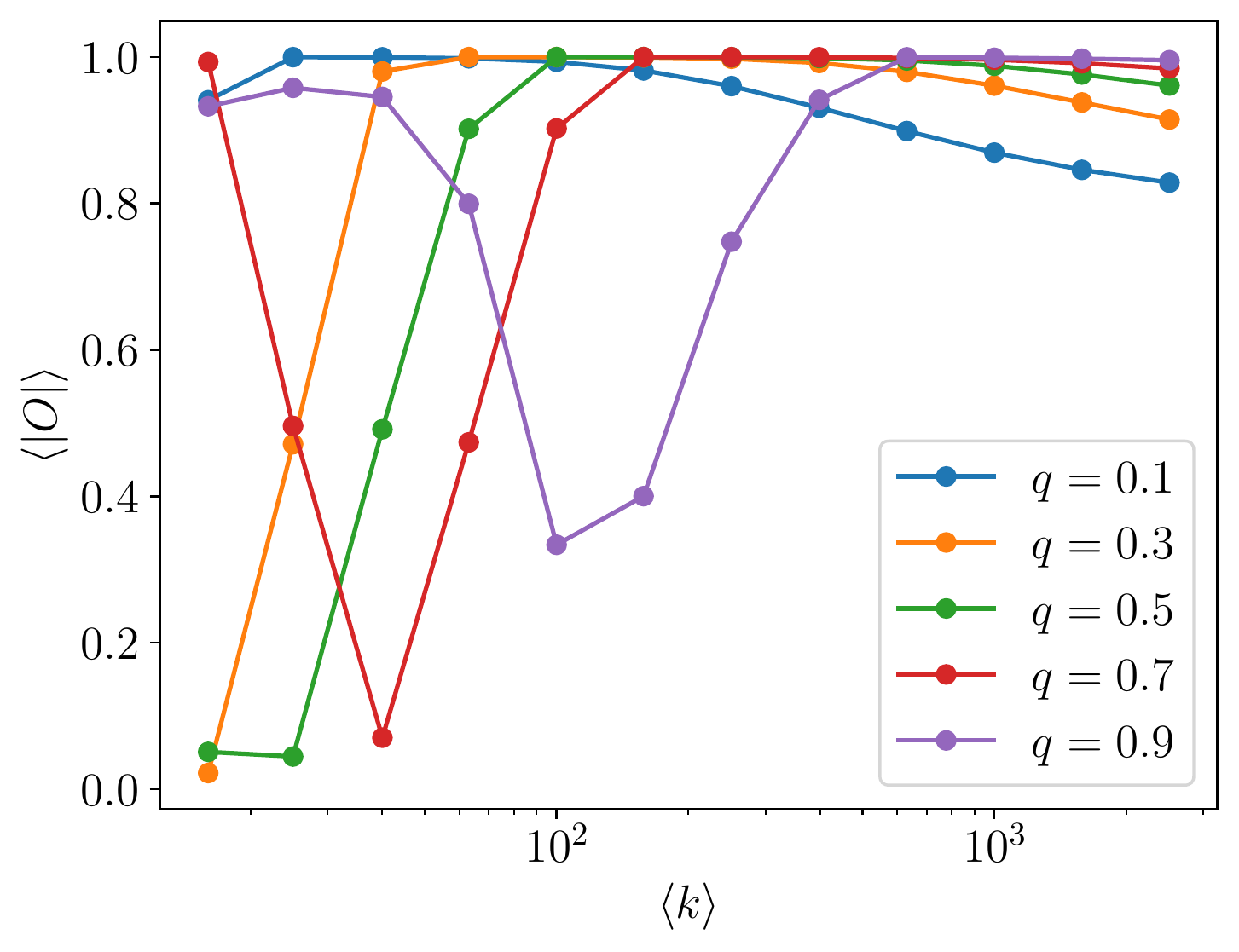}
    \caption{Influence of the connectivity of the network in the stationary state of the average opinion for $f_0 = 0.15$ and different values of $q$. Here we can see that the size of the network has a considerable effect in the outcome of the dynamic, and this is why the approximation of the model is not very accurate.}
    \label{OxK}
\end{figure*}

The relation between connectivity $\langle k \rangle$ and the order parameter is detailed further in Figure \ref{OxK} where the order parameter as a function of average connectivity of the network for different values of $q$ is shown. The order parameter also has a non-monotonic behaviour with $\langle k \rangle$, especially for higher values of $q$. Specifically for $q =0.7$ and $q=0.9$ we see an initial decrease in the order parameter and then a subsequent increase. In general very high connectivity ($\langle k \rangle \sim N$) tends to lead to an ordered state.


\section{Final Remarks}

In this work we studied the effect of memory in the BCS model. In this model the sign of the interactions between the agents evolves alongside the opinions. This introduces a black-listing effect in which if agents have opposing views they start interacting negatively from then on. Conversely, if their interactions are negative and now they agree their interaction can become positive.

This minor change to the model produces vastly different results from the original model. The results are non-monotonic with the probability of the change in the sign of the interaction and with the average degree of the network. Our results indicate that if the network starts with a big fraction of negative interactions there is an optimal value for $q$, that promotes consensus. In the same way we show that for different values of $q$ there is a value of $\langle k \rangle$ that promotes the disordered state. Higher connectivity of the network tend to privilege the ordered state. We also show that for higher values of $q$ the fraction of neutral agents is reduced, to the point of being eliminated for $q=1$.

Memory of the sign of interactions as well as the ability to change these signs to match the opinions have a significant effect in the outcomes of the model. When agents never change their interactions the results are very similar to the original BCS model. As $q$ increases an agent is more likely to change its interaction to match its opinion than changing the opinion itself. This is related to an increased bias against opposing opinions that in the extreme case makes consensus impossible.


\vspace{2.0cm}

\section*{Acknowledgements}

The  authors  acknowledge  financial  support  from  the  Brazilian  funding  agencies  Conselho Nacional de Desenvolvimento Cient\'ifico e Tecnol\'ogico (CNPq, Grants 310893/2020-8 and 302629/2019-0), Coordena\c{c}\~ao de Aperfei\c{c}oamento de Pessoal de N\'ivel Superior (CAPES, Finance code 001) and Funda\c{c}\~ao Carlos Chagas Filho de Amparo \`a Pesquisa do Estado do Rio de Janeiro (FAPERJ, Grant 203.217/2017). We also acknowledge thoughtful remarks by anonymous referees which improved the text.

\vspace{1.0cm}




\appendix
\section{Mean field approximation}

This section describes the mean field approximation for the model. In this approximation all correlations between neighbours and the placement of the negative interactions are ignored. This leads to the discrepancies found between the model and these results.

Considering:
\begin{itemize}
    \item $f$ as the fraction of negative connections;
    \item $g_+$ as the fraction agents with $+1$ opinion;
    \item $g_0$ as the fraction agents with $0$ opinion;
    \item $g_-$ as the fraction agents with $-1$ opinion.
\end{itemize}

\noindent So the rate equations become:

\begin{equation}
    \frac{d f}{dt} = 2q (1-f) g_+ g_- -qf(g_+^2 + g_-^2)
    \label{df}
\end{equation}

\begin{equation}
    \frac{d g_+}{dt} = (1-f)g_+ g_0 + f g_- g_0 - (1-q) g_+ [(1-f)g_- + fg_+]
    \label{dp}
\end{equation}

\begin{equation}
    \frac{d g_-}{dt} = (1-f)g_- g_0 + f g_+ g_0 - (1-q) g_- [(1-f)g_+ + fg_-]
    \label{dm}
\end{equation}

\begin{equation}
    \frac{dg_0}{dt} = (1-q)[2(1-f)g_+ g_- +f(g_+^2 + g_-^2)] - g_0(g_+ + g_-)
    \label{do}
\end{equation}

We still have the following identity $g_+ + g_0 + g_- = 1$, as a consequence we also must have $\frac{dg_+}{dt} +\frac{dg_0}{dt} +\frac{dg_-}{dt} = 0$, this can and was used to check Eqs.~\ref{df}~to~\ref{do}.

We will start by looking into the stationary state results for the fraction $f$. The fraction of negative connections is what drives the ordering or disordering of the system.
From Eq.~(\ref{df}) we can get the stationary state of $f$. For $q \neq 0$ we have

\begin{equation}
    f = \frac{2 g_+ g_-}{{{g_-}^{2}}+g_+ g_-+{{g_+}^{2}}}.
    \label{statf}
\end{equation}

\noindent The first thing to notice is that this result does not explicitly depend on the value of $q$ as long as $q\neq 0$. It is also worth noticing that in the disordered state $g_+ = g_-$, and therefore $f = 2/3$. This result is about double the highest value of $f$ seen in Fig.~\ref{fxp}. The discrepancy seems to come from the fact that a negative connection between agents with different opinions does not behave in the same way as a negative connection between agents with the same opinion.

On the other hand this approximation predicts that in the ordered state when you have either $g_+ = 1$ and $g_- = 0$ or $g_- = 1$ and $g_+ = 0$ we also have $f = 0$. This is in complete agreement with the Monte Carlo simulation results.

Furthermore by substituting Eq.~(\ref{statf}) into Eq.~(\ref{dp}) in the stationary and using $g_0 = 1 - g_+ -g_-$ we get

\begin{equation*}
	\frac{q {{g_-}^{4}}-{{g_-}^{4}}-q g_+ {{g_-}^{3}}-2 g_+ {{g_-}^{3}}+3 q {{g_+}^{2}} {{g_-}^{2}}- 5 {{g_+}^{2}} {{g_-}^{2}}+3 g_+ {{g_-}^{2}}-{{g_+}^{2}} g_--{{g_+}^{4}}+{{g_+}^{3}}}{{{g_-}^{2}}+g_+ g_-+{{g_+}^{2}}}=0
\end{equation*}

\noindent In the disordered state ($g_+ = g_-$) this becomes simply

\begin{equation*}
    \left( q-3\right)  {{g_+}^{2}}+g_+=0
\end{equation*}

\noindent So for $g_+ \neq 0$ we have

\begin{equation}
    g_+ = \frac{1}{3-q}.
    \label{fracs2}
\end{equation}

As $g_0 = 1 - g_+ -g_- = 1 - 2\,g_+$ in the disordered state, from Eq.~(\ref{fracs2}) we have
\begin{equation}
    g_0 = \frac{1-q}{3-q}.
    \label{fracs3}
\end{equation}

The last two results match very well the fractions of opinions found in the simulations as can be seen in Fig.~\ref{hist_p}.


\vspace{1cm}


\begin{thebibliography}{10}
\expandafter\ifx\csname url\endcsname\relax
  \def\url#1{\texttt{#1}}\fi
\expandafter\ifx\csname urlprefix\endcsname\relax\def\urlprefix{URL }\fi
\expandafter\ifx\csname href\endcsname\relax
  \def\href#1#2{#2} \def\path#1{#1}\fi

\bibitem{2012galam}
S.~Galam, Sociophysics, Springer {US}, 2012.
\newblock \href {https://doi.org/10.1007/978-1-4614-2032-3}
  {\path{doi:10.1007/978-1-4614-2032-3}}.

\bibitem{2008galam}
S.~Galam, Sociophysics: A review of galam models, International Journal of
  Modern Physics C 19~(03) (2008) 409--440.
\newblock \href {https://doi.org/10.1142/S0129183108012297}
  {\path{doi:10.1142/S0129183108012297}}.

\bibitem{2014senC}
P.~Sen, B.~K. Chakrabarti, Sociophysics: an introduction, Oxford University
  Press, 2014.

\bibitem{CROKIDAKIS2022127598}
N.~Crokidakis, S.~Galam, After 2018 bolsonaro victory, is a 2022 remake
  feasible?, Physica A: Statistical Mechanics and its Applications 600 (2022)
  127598.

\bibitem{PERALTA2020122475}
A.~F. Peralta, N.~Khalil, R.~Toral, Ordering dynamics in the voter model with
  aging, Physica A: Statistical Mechanics and its Applications 552 (2020)
  122475, tributes of Non-equilibrium Statistical Physics.

\bibitem{Peralta_2020}
A.~F. Peralta, N.~Khalil, R.~Toral, Reduction from non-markovian to markovian
  dynamics: the case of aging in the noisy-voter model, Journal of Statistical
  Mechanics: Theory and Experiment 2020~(2) (2020) 024004.

\bibitem{2009castellanoFL}
C.~Castellano, S.~Fortunato, V.~Loreto, Statistical physics of social dynamics,
  Rev. Mod. Phys. 81 (2009) 591--646.
\newblock \href {https://doi.org/10.1103/RevModPhys.81.591}
  {\path{doi:10.1103/RevModPhys.81.591}}.

\bibitem{2017sirbuLST}
A.~S{\^\i}rbu, V.~Loreto, V.~D. Servedio, F.~Tria, Opinion dynamics: models,
  extensions and external effects, in: Participatory sensing, opinions and
  collective awareness, Springer, 2017, pp. 363--401.

\bibitem{abramiuk2021discontinuous}
A.~Abramiuk-Szurlej, A.~Lipiecki, J.~Paw{\l}owski, K.~Sznajd-Weron,
  Discontinuous phase transitions in the q-voter model with generalized
  anticonformity on random graphs, Scientific Reports 11~(1) (2021) 1--9.

\bibitem{crokidakis2022simple}
N.~Crokidakis, A simple mechanism leading to first-order phase transitions in a
  model of tax evasion, International Journal of Modern Physics C 33~(06)
  (2022) 2250075.

\bibitem{OESTEREICH2020109893}
A.~Oestereich, M.~Pires, S.~{Duarte Queirós}, N.~Crokidakis, Hysteresis and
  disorder-induced order in continuous kinetic-like opinion dynamics in complex
  networks, Chaos, Solitons \& Fractals 137 (2020) 109893.

\bibitem{e21050521}
A.~Abramiuk, J.~Pawłowski, K.~Sznajd-Weron, Is independence necessary for a
  discontinuous phase transition within the q-voter model?, Entropy 21~(5)
  (2019).
\newblock \href {https://doi.org/10.3390/e21050521}
  {\path{doi:10.3390/e21050521}}.

\bibitem{2017gambaroC}
J.~P. Gambaro, N.~Crokidakis, The influence of contrarians in the dynamics of
  opinion formation, Physica A: Statistical Mechanics and its Applications 486
  (2017) 465--472.

\bibitem{castellano2009nonlinear}
C.~Castellano, M.~A. Mu{\~n}oz, R.~Pastor-Satorras, Nonlinear q-voter model,
  Physical Review E 80~(4) (2009) 041129.

\bibitem{2018jedrzejewskiS}
A.~Jedrzejewski, K.~Sznajd-Weron, Impact of memory on opinion dynamics, Physica
  A: Statistical Mechanics and its Applications 505 (2018) 306 -- 315.
\newblock \href {https://doi.org/https://doi.org/10.1016/j.physa.2018.03.077}
  {\path{doi:https://doi.org/10.1016/j.physa.2018.03.077}}.

\bibitem{harris2015random}
R.~J. Harris, Random walkers with extreme value memory: modelling the peak-end
  rule, New Journal of Physics 17~(5) (2015) 053049.

\bibitem{BOSCHI2020124909}
G.~Boschi, C.~Cammarota, R.~Kühn, Opinion dynamics with emergent collective
  memory: A society shaped by its own past, Physica A: Statistical Mechanics
  and its Applications 558 (2020) 124909.

\bibitem{BOSCHI2021125799}
G.~Boschi, C.~Cammarota, R.~Kühn, Opinion dynamics with emergent collective
  memory: The impact of a long and heterogeneous news history, Physica A:
  Statistical Mechanics and its Applications 569 (2021) 125799.

\bibitem{1979lordRL}
C.~G. Lord, L.~Ross, M.~R. Lepper, Biased assimilation and attitude
  polarization: The effects of prior theories on subsequently considered
  evidence., Journal of Personality and Social Psychology 37~(11) (1979)
  2098--2109.
\newblock \href {https://doi.org/10.1037/0022-3514.37.11.2098}
  {\path{doi:10.1037/0022-3514.37.11.2098}}.

\bibitem{1992griffinT}
D.~Griffin, A.~Tversky, The weighing of evidence and the determinants of
  confidence, Cognitive Psychology 24~(3) (1992) 411--435.
\newblock \href {https://doi.org/10.1016/0010-0285(92)90013-r}
  {\path{doi:10.1016/0010-0285(92)90013-r}}.

\bibitem{1998nickerson}
R.~S. Nickerson, Confirmation bias: A ubiquitous phenomenon in many guises,
  Review of General Psychology 2~(2) (1998) 175--220.
\newblock \href {https://doi.org/10.1037/1089-2680.2.2.175}
  {\path{doi:10.1037/1089-2680.2.2.175}}.

\bibitem{2015bessiPVZ}
A.~Bessi, F.~Petroni, M.~D. Vicario, F.~Zollo, A.~Anagnostopoulos, A.~Scala,
  G.~Caldarelli, W.~Quattrociocchi, Viral misinformation, in: Proceedings of
  the 24th International Conference on World Wide Web, {ACM}, 2015.
\newblock \href {https://doi.org/10.1145/2740908.2745939}
  {\path{doi:10.1145/2740908.2745939}}.

\bibitem{2015zolloNVB}
F.~Zollo, P.~K. Novak, M.~D. Vicario, A.~Bessi, I.~Mozeti{\v{c}}, A.~Scala,
  G.~Caldarelli, W.~Quattrociocchi, Emotional dynamics in the age of
  misinformation, {PLOS} {ONE} 10~(9) (2015) e0138740.
\newblock \href {https://doi.org/10.1371/journal.pone.0138740}
  {\path{doi:10.1371/journal.pone.0138740}}.

\bibitem{2016vicarioBZP}
M.~D. Vicario, A.~Bessi, F.~Zollo, F.~Petroni, A.~Scala, G.~Caldarelli, H.~E.
  Stanley, W.~Quattrociocchi, The spreading of misinformation online,
  Proceedings of the National Academy of Sciences 113~(3) (2016) 554--559.
\newblock \href {https://doi.org/10.1073/pnas.1517441113}
  {\path{doi:10.1073/pnas.1517441113}}.

\bibitem{2014allahverdyanG}
A.~E. Allahverdyan, A.~Galstyan, Opinion dynamics with confirmation bias,
  {PLoS} {ONE} 9~(7) (2014) e99557.
\newblock \href {https://doi.org/10.1371/journal.pone.0099557}
  {\path{doi:10.1371/journal.pone.0099557}}.

\bibitem{2021liuWCT}
L.~Liu, X.~Wang, X.~Chen, S.~Tang, Z.~Zheng, Modeling confirmation bias and
  peer pressure in opinion dynamics, Frontiers in Physics 9 (mar 2021).
\newblock \href {https://doi.org/10.3389/fphy.2021.649852}
  {\path{doi:10.3389/fphy.2021.649852}}.

\bibitem{2022anagnostopoulosBCP}
A.~Anagnostopoulos, L.~Becchetti, E.~Cruciani, F.~Pasquale, S.~Rizzo, Biased
  opinion dynamics: when the devil is in the details, Information Sciences 593
  (2022) 49--63.
\newblock \href {https://doi.org/10.1016/j.ins.2022.01.072}
  {\path{doi:10.1016/j.ins.2022.01.072}}.

\bibitem{2000deffuantNAW}
G.~Deffuant, D.~Neau, F.~Amblard, G.~Weisbuch, Mixing beliefs among interacting
  agents, Advances in Complex Systems 03~(01n04) (2000) 87--98.

\bibitem{2019oestereichPC}
A.~L. Oestereich, M.~A. Pires, N.~Crokidakis, Three-state opinion dynamics in
  modular networks, Physical Review E 100~(3) (2019) 032312.
\newblock \href {https://doi.org/10.1103/physreve.100.032312}
  {\path{doi:10.1103/physreve.100.032312}}.

\bibitem{2012biswasCS}
S.~Biswas, A.~Chatterjee, P.~Sen, Disorder induced phase transition in kinetic
  models of opinion dynamics, Physica A: Statistical Mechanics and its
  Applications 391~(11) (2012) 3257 -- 3265.

\bibitem{2010lallouacheCCC}
M.~Lallouache, A.~S. Chakrabarti, A.~Chakraborti, B.~K. Chakrabarti, Opinion
  formation in kinetic exchange models: Spontaneous symmetry-breaking
  transition, Physical Review E 82 (2010) 056112.
\newblock \href {https://doi.org/10.1103/PhysRevE.82.056112}
  {\path{doi:10.1103/PhysRevE.82.056112}}.

\bibitem{RAQUEL2022127825}
M.~Raquel, F.~Lima, T.~Alves, G.~Alves, A.~Macedo-Filho, J.~Plascak,
  Non-equilibrium kinetic biswas–chatterjee–sen model on complex networks,
  Physica A: Statistical Mechanics and its Applications 603 (2022) 127825.

\bibitem{1959erdosR}
P.~Erd{\H{o}}s, A.~R{\'{e}}nyi, On random graphs, i, Publicationes Mathematicae
  (Debrecen) 6 (1959) 290--297.

\end{thebibliography}

\end{document}